\documentclass[10pt]{iopart}
\usepackage{graphicx,float,epsfig,wrapfig,multirow}
\begin{document}

\title[ASTROD, ASTROD I and their gravitational-wave sensitivities]{ASTROD, ASTROD I and their gravitational-wave sensitivities}

\author[W-T Ni and S Shiomi]{Wei-Tou Ni$^{1, 2, 3}$, Sachie Shiomi$^{3}$ and An-Chi Liao$^{3}$}

\address{$^{1}$Purple Mountain Observatory, Chinese Academy of Sciences, Nanjing\\
$^{2}$National Astronomical Observatories, Chinese Academy of Sciences, Beijing\\
$^{3}$Department of Physics, National Tsing Hua University, Hsinchu}

\begin{abstract}
ASTROD (Astrodynamical Space Test of Relativity using Optical Devices) is a mission concept
with three spacecraft --- one near L1/L2 point, one with an inner solar orbit and one with an outer solar
orbit, ranging coherently with one another using lasers to test relativistic gravity, to measure the
solar system and to detect gravitational waves.  ASTROD I with one spacecraft ranging optically
with ground stations is the first step toward the ASTROD mission.  In this paper, we present
the ASTROD I payload and accelerometer requirements, discuss the
gravitational-wave sensitivities for ASTROD and ASTROD I, and compare them with LISA and radio-wave
Doppler-tracking of spacecraft.
\end{abstract}

\ead{wtni@pmo.ac.cn}




\section{Introduction}

A standard implementation of ASTROD is to have two spacecraft in
separate solar orbit carrying a payload of a proof mass, two
telescopes, two 1-2 W lasers, a clock and a drag-free system,
together with a similar L1/L2 spacecraft (Fig.~1)[1, 2]. The three
spacecraft range coherently with one another using lasers.
Simulation with timing error 50 ps and accelerometer error
$10^{-13}$ m/s$^{2}$(Hz)$^{1/2}$ at frequency ${\it f}$ $\sim$
100${\mu}$Hz for 1050 days gives the uncertainties of determining
relativistic parameters $\gamma$ and $\beta$ to be $4.6 \times
10^{-7}$ and $4.0 \times 10^{-7}$, respectively, and that for solar quadrupole
parameter $J_{2}$ to be 1.2$\times 10^{-8}$ [3].  At present,
satellite laser-ranging timing error has reached 5 ps. LISA's
launching goal in 2012 for accelerometer noise requires 3 $\times$
$10^{-15}$ m/s$^{2}$(Hz)$^{1/2}$ at ${\it f}$ $\sim$ 100
${\mu}$Hz [4]. Since ASTROD will be after 2012, a factor of 3
improvement on LISA goal will make 2 orders of magnitude
improvement for acceleration noise in the simulation. These will
put the uncertainties of measuring $\gamma$ and $\beta$ in the
10$^{-9}$ to 10$^{-8}$ range, and that of $J_{2}$ in the
10$^{-10}$ to 10$^{-9}$ range. This demands post-post-Newtonian
ephemeris framework to be established for the analysis and
simulation of data.  Efforts have been started in this direction
by colleagues [5-7].

ASTROD I with one spacecraft interferometric ranging and pulse
ranging with ground stations is the first step for a full ASTROD
mission [8].  The goals are testing relativity with gamma measured
to 10$^{-7}$, measuring solar-system parameters more precisely and
improving the sensitivity for gravitational wave detection using
radio Doppler tracking . The spacecraft is to be launched to
encounter Venus twice to achieve a shorter period for a sooner
measurement of Shapiro time delay (Fig.~2).

\section{ASTROD sensitivity}

The algorisms for unequal arm noise cancellation of Armstrong,
Estabrook \& Tinto [9, 10], and of Dhurandhar, Nayak \& Vinet [11]
can readily be applied to ASTROD. For the 3-spacecraft
configuration as in LISA and as in Fig.~1, they have demonstrated
that one can combine the six measured time series of Doppler
shifts of the one-way laser beams between spacecraft pairs, and
the six measured shifts between adjacent optical benches on each
spacecraft, with suitable time delays, to cancel the laser
frequency fluctuations, and opical-bench vibration noise.  The
achievable strain sensitivity for these combinations is set by the
drag-free proof-mass accelerations inside optical benches, and by
the shot noise at photodetectors.  The implementation of these
alogarisms for ASTROD requires the armlengths known
accurately to about 100 m and the synchronization good to about 300
ns.  Since ASTROD is an astrometry mission, the armlengths and
synchronizations will be measured and implemented to better than 10 ps
accuracy.  Therefore, ASTROD can easily accommodate the noise
cancellation algorisms even when shot noise and acceleration noise
are improved significantly.

\begin{figure}[b]
\begin{center}
\parbox[b]{2.15 in}{\psfig{file = 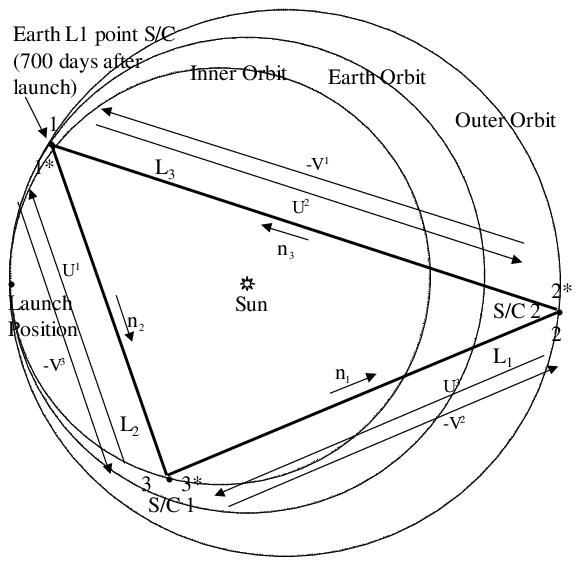, width=2.15 in}}
\parbox[b]{2.95 in}{\psfig{file = 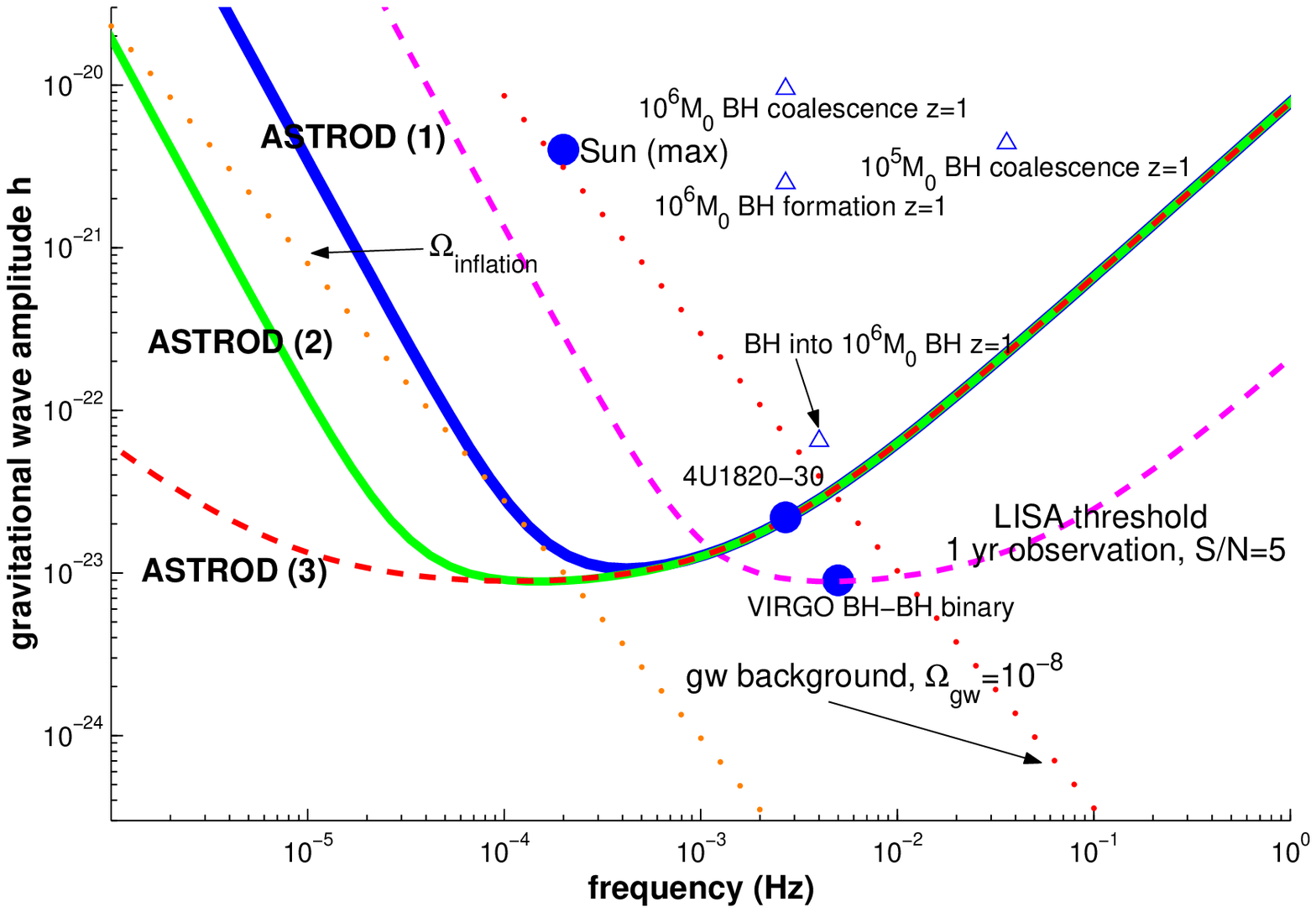, width=2.95 in}}
\begin{flushleft}
\parbox{5.1 in}{\scriptsize \textbf{Figure 1.} (left) A schematic ASTROD
configuration;(right)the gravitational sensitivity curve of ASTROD
and the strength of various sources.}
\end{flushleft}
\end{center}
\end{figure}

For 1-2 W laser emitting power, the shot noise sensitivity
floor in the strain is $10^{-21}$/(Hz)$^{1/2}$ independent of armlength.
With the same power, the ASTROD sensitivity would be
shifted to lower frequency by a factor up to 60 (30 in average) if
other frequency-dependent requirements can be met.  The antenna
response will be shifted correspondingly.  If the accelerometer
noise response can be shifted, i.e., the low-frequency response
becomes better, we may have the ASTROD (2) sensitivity curve of
Fig.~1 as given in [12].  ASTROD has varying armlength and the
sensitivity should be in a band and integration should be
performed. If the LISA accelerometer noise goal is taken, the
sensitivity at low frequency is about 30 times better than LISA as
indicated in ASTROD (1) in Fig.~1, as given by R$\ddot{\textrm{u}}$diger
[13]. Since ASTROD is in a time frame later than LISA, if the
absolute metrological accelerometer / inertial sensor can be
developed, there is a potential to reach ASTROD (3) sensitivity
curve.  With a better lower-frequency resolution, the confusion
limit for LISA should be lower for ASTROD.
The modulations of signals due to orbit motions of ASTROD
spacecraft, especially the radial motions,
will help to distinguish gravity-gradient signals due to solar oscillation from
gravitational-wave signals from outside the solar system [2].

\begin{figure}[b]
\begin{center}
\parbox[b]{2.45 in}{\psfig{file = 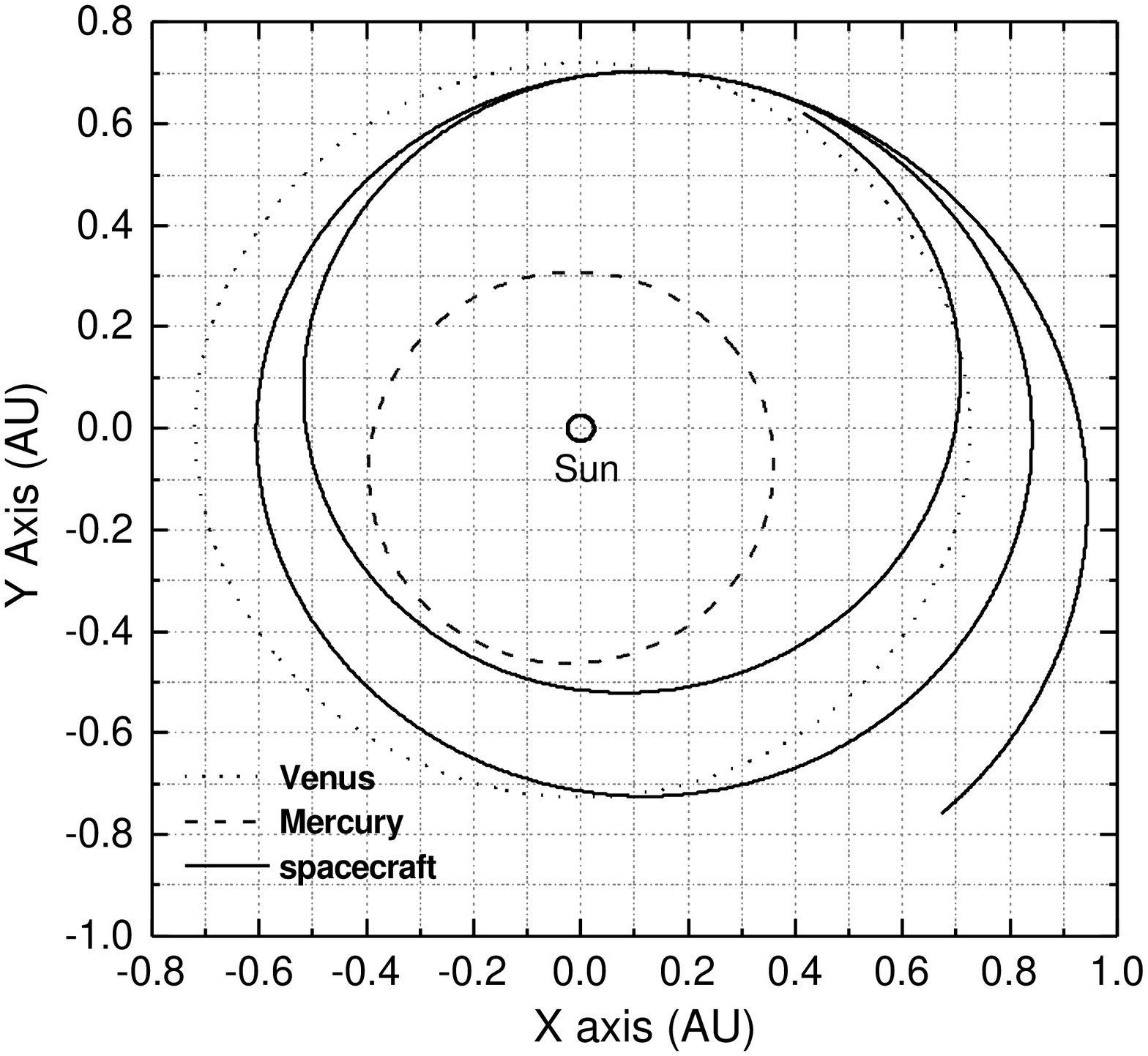, width=2.5 in}}
\parbox[b]{2.65 in}{\psfig{file = 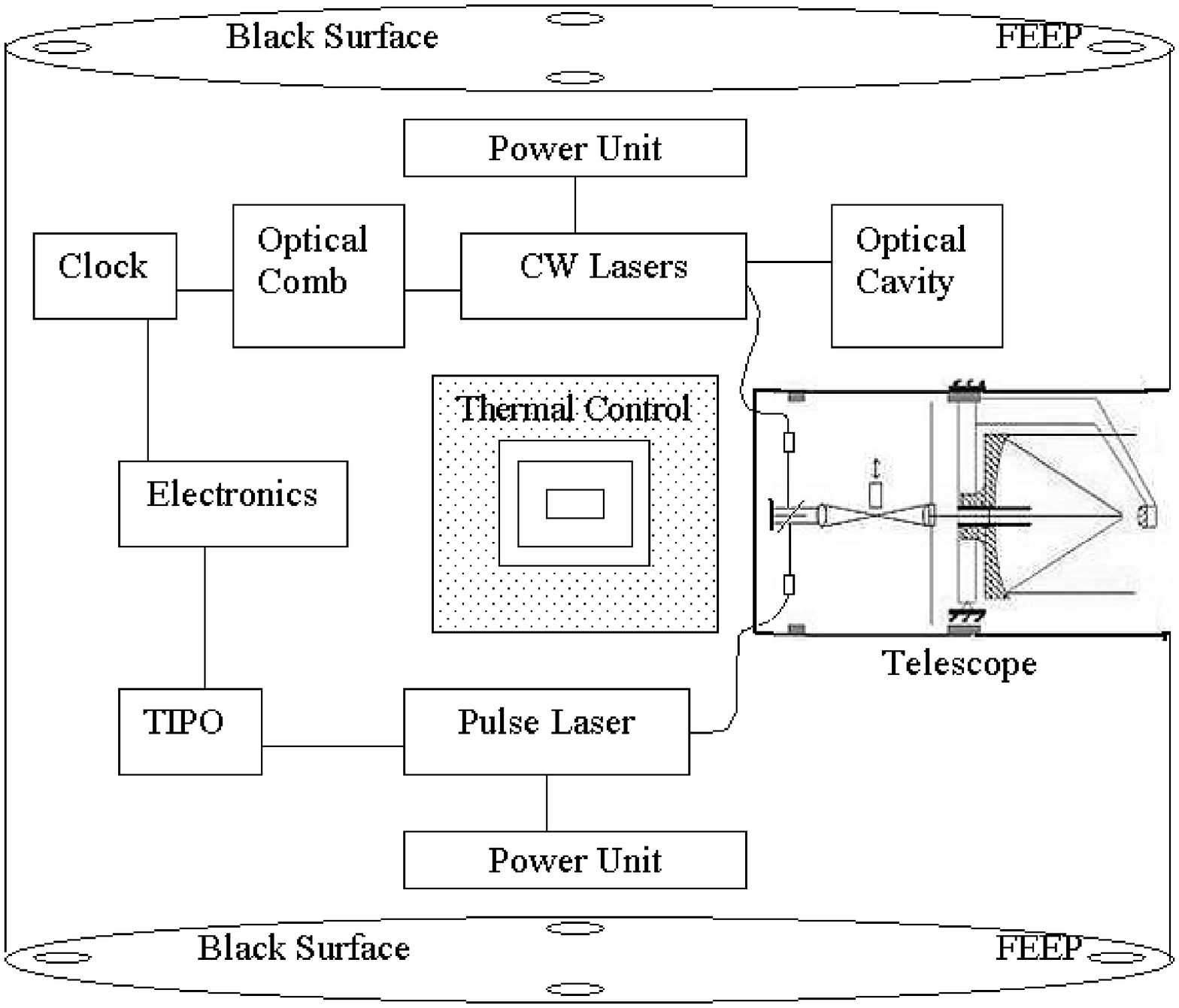, width= 2.73 in}}
\begin{flushleft}
\parbox{5.1 in}{\scriptsize \textbf{Figure 2.} (left) A design orbit of ASTROD I in the
heliocentric ecliptic coordinate system;(right)a schematic diagram
of payload configuration of ASTROD I.}
\end{flushleft}
\end{center}
\end{figure}

\section{ASTROD I payload, requirement and sensitivity}

A schematic payload configuration of ASTROD I is as in Fig.~2 [14].  The cylindrical spacecraft with diameter 2.5 m and height 2 m has its surface covered with solar panels.  In orbit, the cylindrical axis is perpendicular to the orbit plane with the telescope pointing toward the ground laser station.  The effective area to receive sunlight is about 5 m$^2$ and can generate over 500 W of power.  The total mass of spacecraft is 300-350 kg.  That of payload is 100-120 kg with science data rate 500 bps.
The crucial technologies include (i) 100 fW weaklight phase locking [15]; (ii) design and development of sunlight shield [8]; (iii) design and development of drag-free system. The requirements for the drag-free system are less stringent than LISA.
In Table 1 we provide a summary of the parameter values that meet the
accelerometer requirements for ASTROD I, in comparison with LISA. We use the notations and the LISA
values of Schumaker [16].  The
listed values are for the frequency at 0.1 mHz. The
parameter values are largely relaxed for ASTROD I. The same values satisfy requirements for higher-frequency
range, where frequency-dependent contributions to the
acceleration noise become smaller.

\begin{table}[t]
\noindent
 \begin{center}
 \fontsize{8}{10pt} \selectfont
\parbox{13 cm}{\scriptsize \hspace{2 pt} \textbf{Table 1.} Parameter values that meet ASTROD I accelerometer requirements, as compared
with LISA.}
 \begin{tabular}{@{}lll}
  \br
 $f = (2 \pi)^{-1} \omega =$ 0.1 $\textrm{mHz}$  & \textbf{ASTROD I} & \textbf{LISA} \\
  \mr
  \textit{Acceleration noise goal at 0.1 $\textrm{mHz}$:}
    $A_p$ [$\textrm{m s}^{-2}\textrm{ Hz}^{-1/2}$] & $10^{-13}$ & $3 \times 10^{-15}$ \\
  \mr
  \textit{Parameter values} & & \\
    \textbf{Proof Mass (PM)} \\
  Magnetic susceptibility: $\chi_m$ & $10^{-5}$ & $10^{-6}$ \\
  Maximum charge build-up: $q$ [$\mathrm{C}$] & $10^{-12}$ & $10^{-13}$ \\
  Residual gas pressure: $P$ [$\mathrm{Pa}$] & $10^{-5}$ & $10^{-6}$ \\
  Fluctuation of temperature difference across & \multirow{2}{*}{1.4 $\times$ $10^{-3}$} & \multirow{2}{*}{2.2 $\times$ $10^{-5}$} \\
  \hspace{1 pt} PM and housing: $\delta T_d$ [$\textrm{K } \textrm{Hz}^{-1/2}$] (at 0.1 $\textrm{mHz}$) & & \\
  \textbf{Spacecraft (SC)} \\
  Thruster noise [$\mu\textrm{N } \textrm{Hz}^{-1/2}$] (at 0.1 $\textrm{mHz}$) & 0.5 & 0.1 \\
  Fluctuation of temperature in SC: $\delta T_{sc}$ [$\textrm{K } \textrm{Hz}^{-1/2}$]
  (at 0.1 $\textrm{mHz}$) & 0.4 & 0.004 \\
  \textbf{Capacitive sensing}\\
  Voltage difference between average voltage across opposite & \multirow{2}{*}{1} & \multirow{2}{*}{0.1}\\
  \hspace{1 pt} faces and voltage to ground: $V_{0g}$ [$\mathrm{V}$] &  &  \\
  Fluctuation of voltage difference across opposite faces: & \multirow{2}{*}{$10^{-4}$} & \multirow{2}{*}{$10^{-5}$} \\
  \hspace{1 pt} $\delta V_d$ [$\textrm{V Hz}^{-1/2}$] (at 0.1 $\textrm{mHz}$) &  & \\
  Asymmetry in gap across opposite sides of PM: $\Delta d$ [$\mu \textrm{m}$] & 10 & 1 \\
  \textbf{Laser power} \\
  Fluctuation of laser power: $\delta I$ [$\textrm{W Hz}^{-1/2}$] (at 0.1 $\textrm{mHz}$) & 2 $\times$
  $10^{-6}$ & 2 $\times$ $10^{-8}$ \\
    \br
\end{tabular}
\end{center}
\end{table}

A typical design orbit for August 4, 2010 launch is shown on the
left of Fig~2: two encounters with Venus to swing the spacecraft to
the other side of the Sun to conduct Shapiro time delay
measurement and to measure Venus multipole moments [17]. Assuming
a 10 ps timing accuracy and $10^{-13}$ m/s$^{2}$(Hz)$^{1/2}$ (at
${\it f}$ $\sim {100\mu}$Hz) inertial sensor/accelerometer noise,
our simulation of the accuracy for determining the relativistic
parameters $\gamma$ and $\beta$, and the solar quadrupole
parameter J$_2$  gives 10$^{-7}$, 10$^{-7}$ and 10$^{-8}$, respectivly, for their uncertainties [17].

Like the radio Doppler tracking of spacecraft [18], ASTROD I also
has sensitivity to low-frequency gravitational-wave.
 Clock noise and the propagation noise are the dominant noise sources. Clock noise
is the dominant instrumental noise.  As clock stability improves,
this noise becomes smaller.  The propagation noise is due to fluctuations in the index of refraction of the troposphere, ionosphere and the interplanetary solar plasma.  The fluctuations due to ionosphere and interplanetary plasma are not important for laser ranging.  Tropospherical effect can be substracted to a large extent by 2-color (2-wavelength) laser ranging or
by using artificial stars.  The ASTROD I spacecraft noise is very small and negligible.  With these improvements, ASTROD I
will have a couple of times to several times better sensitivity to gravitational-wave for the same spacecraft-Earth configuration. Detailed analysis will be presented in a forthcoming publication.

We thank Foundation of Minor Planets of Purple Mountain
Observatory and National Science Council [Grant no. NSC
92-2112-M-007-042] for financial supports.

\References

\bibitem[1] {} Bec-Borsenberger A {\it et al} 2000 {\it ASTROD} ESA F2/F3 Mission Proposal; and references therein

\bibitem[2] {} Ni W-T 2002 {\it Int. J. Mod. Phys.} {\bf D11} 947; and references therein

\bibitem[3] {} Chiou D-W, and Ni W-T 2000 Orbit Simulation for the Determination of Relativistic and Solar-System Parameters for the ASTROD Space Mission, 33rd COSPAR Scientific Assembly, Warsaw, 16-23 July, 2000; Chiou D-W, and Ni W-T 2000 {\it Advances in Space Research} {\bf 25} 1259

\bibitem[4] {} LISA: System and Technology Study Report 2000 ESA document ESA-SCI (2000) 11,
July 2000, revised as
ftp://ftp.rzg.mpg.de/pub/grav/lisa/sts/sts\_1.05.pdf; and
references therein

\bibitem[5] {} Huang T, and Tao J 2002 {\it  Int. J. Mod. Phys.} {\bf D11} 1011

\bibitem[6] {} Li G, and Zhao H 2002 {\it  Int. J. Mod. Phys.} {\bf D11} 1021

\bibitem[7] {} Xu C-M, and Wu X-J 2003 {\it  Chin. Phys. Lett.} {\bf 20} 195

\bibitem[8] {} Ni W-T {\it et al} 2002 {\it Int. J. Mod. Phys.} {\bf D11} 1035; and references therein

\bibitem[9] {} Tinto M, Estabrook F B and Armstrong J W 2002 {\it Phys. Rev.} {\bf D65} 082003; and references therein

\bibitem[10] {} Prince T A, Tinto M, Larson S L and Armstrong J W 2002 {\it Phys. Rev.} {\bf D66} 122002

\bibitem[11] {Dhurandhar} Dhurandhar S V, Nayak K R,Vinet J-Y 2002 {\it Phys. Rev.} {\bf D65} 102002

\bibitem[12] {} Ni W-T 1997 {\it Gravitational Wave Detection} 117 eds. Tsubono K {\it et al} (Tokyo: UAP)

\bibitem[13] {} R$\ddot{\textrm{u}}$diger A 2002 {\it Int. J. Mod. Phys.} {\bf D11} 963

\bibitem[14] {} {\it ASTROD I Pre-Phase A Report} 2003 in preparation

\bibitem[15] {} Liao A-C, Ni W-T, and Shy J-T 2002 {\it Int. J. Mod. Phys.} {\bf D11} 1075

\bibitem[16] {} Schumaker B L 2003 {\it Class. Quantum Grav.} {\bf 20} S239; and references therein

\bibitem[17] {} Tang C J {\it et al} 2003 {\it Orbit Design for the ASTROD I} paper in preparation

\bibitem[18] {} Tinto M 2002 {\it Class. Quantum Grav.} {\bf 19} 1767; and references therein

\endrefs

\end{document}